\documentclass[fleqn,usenatbib,useAMS]{mnras}
 \usepackage{graphicx}
 \usepackage{amsmath}
 \usepackage{amssymb}
 \usepackage[T1]{fontenc}
 \usepackage{ae,aecompl}
 \usepackage{txfonts}
 \usepackage{enumerate}

 \title[Asteroid 2014 YX$_{49}$: a Trojan of Uranus]
       {Asteroid 2014 YX$_\mathbf{49}$: a large transient Trojan of Uranus}

 \author[C. de la Fuente Marcos and R. de la Fuente Marcos]
        {C.~de~la~Fuente~Marcos\thanks{E-mail: nbplanet@ucm.es}
         and
         R. de la Fuente Marcos \\
         Universidad Complutense de Madrid,
         Ciudad Universitaria, E-28040 Madrid, Spain}
 \date{Accepted 2017 January 19. 
       Received 2017 January 17; 
       in original form 2016 December 9}
 \pubyear{2017}
 
\begin{document}
  \label{firstpage}
  \pagerange{\pageref{firstpage}--\pageref{lastpage}}
  \maketitle

  \begin{abstract}
     In the outer Solar system, primordial Trojan asteroids may have remained 
     dynamically stable for billions of years. Several thousands of them 
     accompany Jupiter in its journey around the Sun and a similarly large 
     population may be hosted by Neptune. In addition, recently captured or 
     transient Jovian and Neptunian Trojans are not uncommon. In contrast, no 
     Trojans of Saturn have been found yet and just one Uranian Trojan is 
     known, 2011~QF$_{99}$. Here, we discuss the identification of a second 
     Trojan of Uranus: 2014~YX$_{49}$. Like 2011~QF$_{99}$, 2014~YX$_{49}$ is 
     a transient L$_4$ Trojan although it orbits at higher inclination 
     (25\fdg55 versus 10\fdg83), is larger (absolute magnitude of 8.5 versus 
     9.7) and its libration period is slightly shorter (5.1 versus 5.9~kyr); 
     contrary to 2011~QF$_{99}$, its discovery was not the result of a 
     targeted survey. It is less stable than 2011~QF$_{99}$; our extensive 
     $N$-body simulations show that 2014~YX$_{49}$ may have been following a 
     tadpole trajectory ahead of Uranus for about 60~kyr and it can continue 
     doing so for another 80~kyr. Our analysis suggests that it may remain as 
     co-orbital for nearly 1~Myr. As in the case of 2011~QF$_{99}$, the 
     long-term stability of 2014~YX$_{49}$ is controlled by Jupiter and 
     Neptune, but it is currently trapped in the 7:20 mean motion resonance 
     with Saturn. Consistently, the dynamical mechanism leading to the 
     capture into and the ejection from the Trojan state involves ephemeral 
     multibody mean motion resonances.  
  \end{abstract}


  \section{Introduction}
     Among the giant planets, Saturn and Uranus lack a significant present-day population of small bodies subjected to their respective 1:1 
     mean motion or co-orbital resonances (quasi-satellite, Trojan, horseshoe or combinations of these three elementary resonant states). 
     This is particularly true in the case of Trojans or minor bodies librating 60\degr{} ahead (L$_4$ Lagrangian point) or behind (L$_5$) a 
     host planet in its orbit. 
     \hfil\par
     Trojans describe a shifting path in the shape of a tadpole around the associated Lagrangian point when seen in a frame of reference 
     centred at the Sun and rotating with the host planet (for further details, see e.g. Murray \& Dermott 1999). Jupiter is accompanied by 
     many thousands of them (see e.g. Milani 1993; Jewitt, Trujillo \& Luu 2000; Morbidelli et al. 2005; Yoshida \& Nakamura 2005; Robutel 
     \& Gabern 2006; Robutel \& Bodossian 2009; Grav et al. 2011; Nesvorn\'y, Vokrouhlick\'y \& Morbidelli 2013) and Neptune may have a 
     similarly large population including both primordial or long-term stable Trojans (e.g. Kortenkamp, Malhotra \& Michtchenko 2004; 
     Nesvorn\'y \& Vokrouhlick\'y 2009; Sheppard \& Trujillo 2006, 2010) and transient or recently captured ones (e.g. de la Fuente Marcos 
     \& de la Fuente Marcos 2012b,c). In contrast to the cases of Jupiter and Neptune, any existing Trojans of Saturn still remain to be 
     detected and the first Trojan of Uranus ---a transient one, 2011~QF$_{99}$--- was announced in 2013 (Alexandersen et al. 2013a). This 
     long-sought first Uranian Trojan was found within the context of a carefully planned survey (Alexandersen et al. 2013b).  
     \hfil\par
     Alexandersen et al. (2013b) have shown that 2011~QF$_{99}$ may stay as L$_4$ Trojan of Uranus for 100 kyr to 1 Myr; prior to and/or 
     after leaving its current tadpole path, 2011~QF$_{99}$ may have experienced other co-orbital states. Unlike 2011~QF$_{99}$, Jovian or 
     Neptunian primordial Trojans may have remained as such for the age of the Solar system or over 4.5 Gyr. In total, 2011~QF$_{99}$ may 
     spend about 3 Myr trapped in a co-orbital resonance with Uranus. Asteroid 2011~QF$_{99}$ could be a former member of the Centaur 
     population and it may return to it after being scattered away at the end of its current co-orbital episode. 
     \hfil\par
     The present-day dynamical status as Uranian Trojan of 2011~QF$_{99}$ was further confirmed by de la Fuente Marcos \& de la Fuente 
     Marcos (2014) who showed that a three-body mean motion resonance is responsible for both its insertion into Uranus' co-orbital region 
     and its eventual escape from there. The critical role of three-body resonances as source of instability for Uranian Trojans was first 
     discussed by Marzari, Tricarico \& Scholl (2003). Here, we identify a second Trojan of Uranus, 2014~YX$_{49}$, and explore numerically 
     its past, current and future dynamical evolution. This paper is organized as follows. In Section 2, we show the data available on 
     2014 YX$_{49}$ and discuss our methodology. The orbital evolution of 2014 YX$_{49}$ is studied in Section 3, where its current 
     dynamical status as Uranian Trojan is confirmed statistically. In Section 4, we discuss our results within the context of our present 
     knowledge of the population of Uranian co-orbitals. Our conclusions are summarized in Section 5.

  \section{Asteroid 2014 YX$_\mathbf{49}$: data and methodology}
     Asteroid 2014 YX$_{49}$ was discovered on 2014 December 26 by B. Gibson, T. Goggia, N. Primak, A. Schultz and M. Willman (Gibson et al. 
     2016)\footnote{http://www.minorplanetcenter.net/iau/mpec/K16/K16O10.html} observing with the 1.8-m Ritchey-Chretien telescope of the 
     Pan-STARRS Project (Kaiser et al. 2004) from Haleakala. At discovery time, its $w$-magnitude was 21.4, its right ascension was 
     7$^{\rm h}$ 29$^{\rm m}$ 22\fs659, and its declination was +30\degr{} 24\arcmin{} 00\farcs37 at an heliocentric distance of 18.398~au. 
     As the primary mission of the Pan-STARRS astronomical survey is to detect incoming near-Earth objects not minor bodies moving within 
     the outer Solar system, the discovery of 2014 YX$_{49}$ can be considered of serendipitous nature.
     \hfil\par
     The discovery was made public on 2016 July 16 (Gibson et al. 2016) and during the following month a relatively large number of 
     precovery images from other surveys were identified, which translated into a rapid improvement of its orbital determination. The 
     solution for the orbit of 2014 YX$_{49}$ currently available (as of 2017 January 9) from Jet Propulsion Laboratory's (JPL) Small-Body 
     Database\footnote{http://ssd.jpl.nasa.gov/sbdb.cgi} is statistically robust (see Table \ref{elements}) as it is based on 70 
     observations spanning a data-arc of 4\,876 days or 13.35 yr, from 2001 September 22 to 2015 January 28. Formally classified dynamically 
     as a Centaur, it has a value of the semimajor axis $a$ = 19.13~au, and moves in an eccentric, $e$ = 0.28, and rather inclined path, $i$ 
     = 25\fdg55. It is a comparatively large object with $H$ = 8.5~mag, which implies a diameter in the range 40--120 km for an assumed 
     albedo of 0.50--0.05. Its period of revolution around the Sun, 83.7$\pm$1.3 yr at present, matches well that of Uranus and this fact 
     makes it a strong candidate to moving co-orbital with this giant planet. 
%
%
         \begin{table}
          \fontsize{8}{11pt}\selectfont
          \tabcolsep 0.30truecm
          \caption{Heliocentric Keplerian orbital elements of 2014~YX$_{49}$ utilized in this paper. This orbital determination is based on 
                   70 observations spanning a data-arc of 4\,876 days or 13.35 yr, from 2001 September 22 to 2015 January 28. In addition to
                   the nominal value of each parameter, the 1$\sigma$ uncertainty is provided as well. The solution is computed at epoch JD 
                   2457800.5, which corresponds to 0:00 UT on 2017 February 16 (J2000.0 ecliptic and equinox). This reference epoch 
                   coincides with the $t = 0$ instant in the figures. Source: JPL's Small-Body Database.
                  }
          \begin{tabular}{lcc}
           \hline
            Semimajor axis, $a$ (au)                          & = & 19.1304$\pm$0.0005 \\
            Eccentricity, $e$                                 & = & 0.276539$\pm$0.000013 \\
            Inclination, $i$ (\degr)                          & = & 25.55097$\pm$0.00002 \\
            Longitude of the ascending node, $\Omega$ (\degr) & = & 91.44425$\pm$0.00003 \\
            Argument of perihelion, $\omega$ (\degr)          & = & 280.584$\pm$0.003 \\
            Mean anomaly, $M$ (\degr)                         & = & 75.587$\pm$0.005 \\
            Perihelion, $q$ (au)                              & = & 13.84012$\pm$0.00015 \\
            Aphelion, $Q$ (au)                                & = & 24.4207$\pm$0.0007 \\
            Absolute magnitude, $H$ (mag)                     & = & 8.5 \\
           \hline
          \end{tabular}
          \label{elements}
         \end{table}
%
%
     \hfil\par
     The value of the relative semimajor axis of 2014 YX$_{49}$ with respect to Uranus, $|a - a_{\rm U}| = 0.00012$~au, is lower than those 
     of any other previously documented Uranian co-orbitals ---(83982) Crantor (0.304~au), 2011~QF$_{99}$ (0.079~au) and (472651) 
     2015~DB$_{216}$ (0.087~au)--- or candidates ---1999~HD$_{12}$ (0.210~au), 2002~VG$_{131}$ (0.026~au) and 2010~EU$_{65}$ (0.265~au). 
     Uranus' co-orbital region goes approximately from 19.0 to 19.4 au, although co-orbitals with $a$ in the range 19.1--19.2 au are far 
     less unstable (de la Fuente Marcos \& de la Fuente Marcos 2015). The orbit of 2014~YX$_{49}$ in Table \ref{elements} indicates that it 
     can only experience close encounters with Uranus because its perihelion distance is well beyond the aphelion of Saturn and its aphelion 
     distance is far shorter than the perihelion of Neptune; this property suggests that 2014~YX$_{49}$ may be as dynamically stable as 
     2011~QF$_{99}$ or 472651. 
     \hfil\par
     The path currently followed by 2014~YX$_{49}$ is more eccentric than that of 2011~QF$_{99}$ (0.277 versus 0.178) and also more inclined
     (25\fdg55 versus 10\fdg83). Although the values of their respective arguments of perihelion are separated by just 7\fdg5, their 
     longitudes of the ascending node are over 131\degr{} apart. As for a comparison between the orbits of 2014~YX$_{49}$ and 472651, the 
     former moves in a less eccentric (0.277 versus 0.326) and also less inclined (25\fdg55 versus 37\fdg70) path. In terms of its 
     present-day orbit, the dynamical behaviour of 2014~YX$_{49}$ could be intermediate between those of 2011~QF$_{99}$ and 472651.
     \hfil\par
     The actual dynamical status of prospective co-orbital bodies is not assessed directly from their present-day orbits but from the 
     analysis of a relevant set of numerical integrations. Such analysis focuses on the study of the behaviour of a critical angle that, in 
     the case of Uranus' co-orbital candidates, is the relative mean longitude $\lambda_{\rm r} = \lambda - \lambda_{\rm U}$, where 
     $\lambda$ and $\lambda_{\rm U}$ are the mean longitudes of the object and Uranus, respectively; $\lambda$ = $M$ + $\Omega$ + $\omega$, 
     where $M$ is the mean anomaly, $\Omega$ is the longitude of the ascending node, and $\omega$ is the argument of perihelion (see e.g. 
     Murray \& Dermott 1999). 
     \hfil\par
     For non-co-orbital or passing bodies $\lambda_{\rm r}\in(0, 360)\degr$, with all the values being equally probable; this behaviour is 
     known as circulation. True co-orbitals exhibit an oscillatory evolution over time of the value of $\lambda_{\rm r}$; this behaviour is 
     known as libration. For a L$_4$ Trojan the libration is around +60\degr, for a L$_5$ Trojan the libration is around $-$60\degr{} or 
     300\degr{} (see e.g. Murray \& Dermott 1999), although the oscillation centre could be significantly shifted from the typical 
     equilateral location in the case of eccentric orbits (Namouni, Christou \& Murray 1999; Namouni \& Murray 2000). The current value of 
     $\lambda_{\rm r}$ for 2014~YX$_{49}$ is $\sim$61\degr{}, which suggests that 2014~YX$_{49}$ could be indeed a present-day Uranian 
     Trojan, but this must be confirmed using $N$-body simulations (for 2011~QF$_{99}$, a Trojan, is $\sim$47\degr{} and $\sim$170\degr{} 
     for 472651, a horseshoe librator).
     \hfil\par
     As in our previous studies on the dynamics of Crantor (de la Fuente Marcos \& de la Fuente Marcos 2013), 2011~QF$_{99}$ (de la Fuente 
     Marcos \& de la Fuente Marcos 2014) and 472651 (de la Fuente Marcos \& de la Fuente Marcos 2015), we use direct $N$-body integrations 
     performed with a modified version of a code written by Aarseth (2003) ---the standard version of this software is publicly available 
     from the IoA website\footnote{http://www.ast.cam.ac.uk/$\sim$sverre/web/pages/nbody.htm}--- that implements the Hermite integration 
     scheme described by Makino (1991). The performance of this code, when applied to Solar system numerical investigations, has been 
     studied thoroughly (for further details, see de la Fuente Marcos \& de la Fuente Marcos 2012a). The physical model is the same one used 
     in the works on Crantor, 2011~QF$_{99}$ and 472651 cited above and includes the perturbations by the eight major planets, the Moon, the 
     barycentre of the Pluto-Charon system, and the three largest asteroids. Initial positions and velocities are based on the DE405 
     planetary orbital ephemerides (Standish 1998) referred to the barycentre of the Solar system and to the epoch JD TDB 2457800.5 
     (2017-February-16.0), which is the $t$ = 0 instant in our simulations. They have been provided by JPL's online Solar system data 
     service\footnote{http://ssd.jpl.nasa.gov/?planet\_pos} (Giorgini et al. 1996). 

  \section{Asteroid 2014 YX$_\mathbf{49}$: orbital evolution}
     In this section, we show the results of simulations that use the nominal orbit in Table \ref{elements} and those of additional control 
     calculations based on sets of orbital elements obtained from the nominal ones as described in de la Fuente Marcos \& de la Fuente 
     Marcos (2013, 2014, 2015). The main set of numerical integrations explores the dynamical evolution of 2014~YX$_{49}$ for 0.6 Myr 
     forward and backwards in time. Another set of shorter simulations aimed at characterizing the short-term stability of 2014~YX$_{49}$ 
     applies the implementation of the Monte Carlo using the Covariance Matrix (MCCM) method discussed by de la Fuente Marcos \& de la 
     Fuente Marcos (2015).

     \subsection{Current dynamical state}
\begin{figure}
        \centering
         \includegraphics[width=\linewidth]{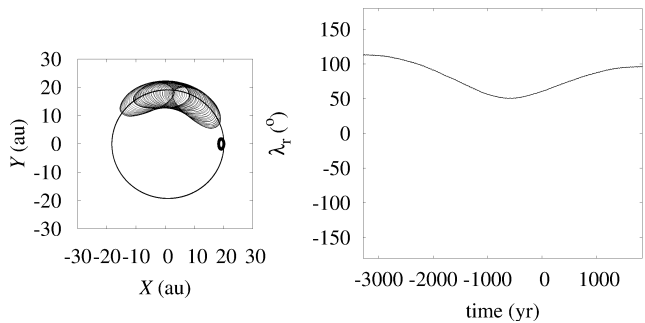}
         \caption{One oscillation of the Trojan motion of 2014~YX$_{49}$ during the time interval ($-$3\,286,~1\,840) yr projected on to the 
                  ecliptic plane in a coordinate system centred at the Sun and rotating with Uranus (left-hand panel). The path followed by
                  Uranus and its position are plotted as well; in this frame of reference, and as a result of the non-negligible value of 
                  its orbital eccentricity, Uranus follows a small ellipse. The values of the resonant angle, $\lambda_{\rm r}$, for one 
                  libration period are also shown (right-hand panel). Nominal orbital solution as in Table \ref{elements} although the 
                  behaviour observed is virtually identical across control orbits within the displayed time frame.
                 }
         \label{orbit}
      \end{figure}

        Fig. \ref{orbit}, left-hand panel, displays the motion of 2014~YX$_{49}$ as characterized by the nominal orbit shown in 
        Table~\ref{elements} over the time range ($-$3\,286,~1\,840)~yr projected on to the ecliptic plane in a coordinate system centred at 
        the Sun and rotating with Uranus. This figure is equivalent to fig. 1 for 2011~QF$_{99}$ in Alexandersen et al. (2013b) or fig. 1 in 
        de la Fuente Marcos \& de la Fuente Marcos (2014). Only one oscillation of the tadpole motion of 2014~YX$_{49}$ is displayed (see 
        the evolution of $\lambda_{\rm r}$ in Fig. \ref{orbit}, right-hand panel). All the control orbits (several hundreds) show virtually 
        the same behaviour in the neighbourhood of $t$ = 0. Alexandersen et al. (2013b) have found a libration period for 2011~QF$_{99}$ of 
        5.9~kyr. Our calculations indicate that the libration period of 2014~YX$_{49}$ is slightly shorter, of about 5.1~kyr. 
        Fig.~\ref{orbit} confirms 2014~YX$_{49}$ as the second Trojan of Uranus; this conclusion is robust because the results of the 
        evolution of a large set of statistically consistent short integrations (lasting $\pm20$~kyr) are virtually identical to each other.
        \hfil\par
%
%
\begin{figure}
        \centering
         \includegraphics[width=\linewidth]{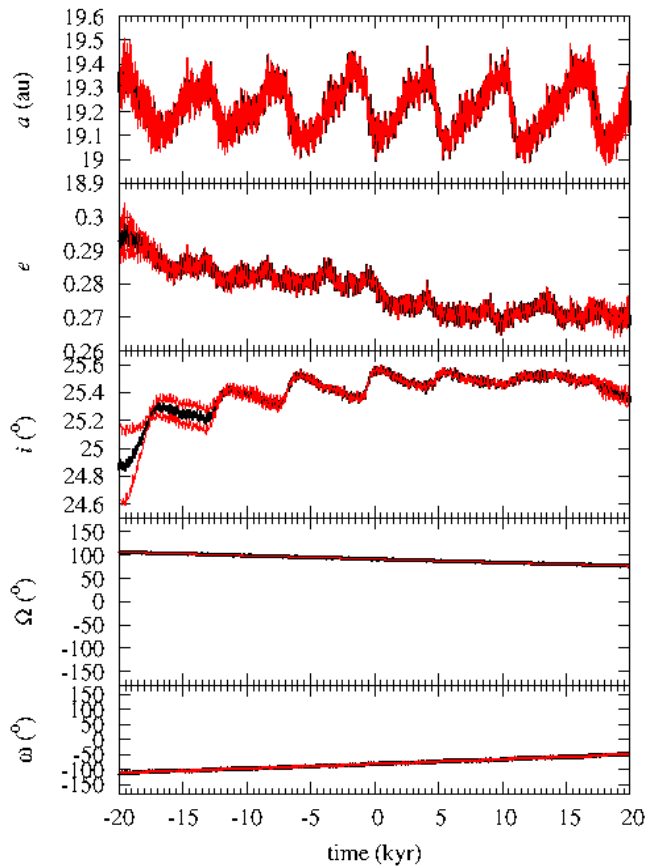}
         \caption{Variation of the orbital elements $a$ (top panel), $e$ (second to top panel), $i$ (middle panel), $\Omega$ (second to 
                  bottom panel), and $\omega$ (bottom panel) of 2014~YX$_{49}$ over time ($\pm$20\,000~yr). The average evolution of 250 
                  control orbits is shown as a thick black curve; the corresponding ranges in the values of the orbital parameters are 
                  displayed as thin red curves. The control orbits utilized to generate the initial conditions for this set of simulations 
                  have been calculated as described in section 3 of de la Fuente Marcos \& de la Fuente Marcos (2015), using the covariance 
                  matrix. 
                 }
         \label{disper}
      \end{figure}

%
%
        Fig. \ref{disper} shows the variation of the orbital elements $a$, $e$, $i$, $\Omega$, and $\omega$ of 2014~YX$_{49}$ over time. 
        This figure displays the average evolution (black curve) of 250 control orbits obtained via MCCM and its associated dispersion (red 
        curves). These control orbits have been computed as described in section 3 of de la Fuente Marcos \& de la Fuente Marcos (2015). The 
        covariance matrix used to perform these calculations was provided by JPL's Small-Body Database. The overall level of stability that 
        characterizes the orbital solution currently available for 2014~YX$_{49}$ can be evaluated quantitatively by computing its 
        associated Lyapunov time or time-scale for exponential divergence of orbits that are started arbitrarily close to each other. Our 
        calculations show that the Lyapunov time is somewhat different for forward ($\sim$20\,000~yr) and backwards ($\sim$9\,000~yr) 
        integrations; this is relatively common in the case of transient co-orbitals. For instance, the Lyapunov time of 2011~QF$_{99}$ is 
        longer for backwards integrations. In the short term, 2014~YX$_{49}$ is clearly more stable than (472651) 2015~DB$_{216}$ (see fig. 
        1 in de la Fuente Marcos \& de la Fuente Marcos 2015); asteroid 472651 is a Uranian asymmetric horseshoe librator. 
%
%
%
%

     \subsection{Past and future orbital evolution}
\begin{figure*}
        \centering
         \includegraphics[width=\linewidth]{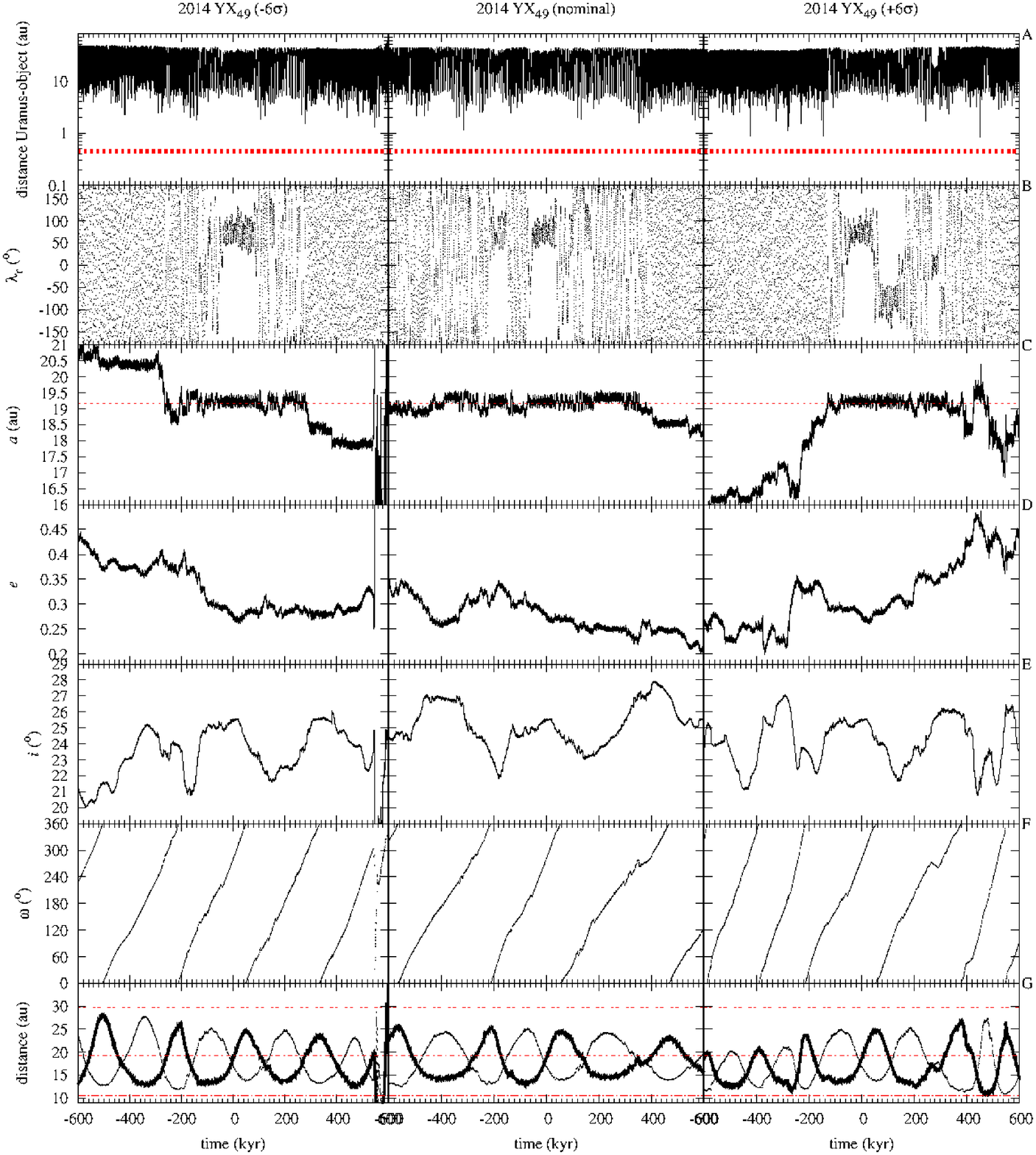}
         \caption{Past and future evolution of various relevant parameters for the nominal or reference orbit of 2014~YX$_{49}$ in 
                  Table~\ref{elements} (central panels) and two illustrative orbits that are most different from the reference one (see the 
                  text for details). The variation of the distance from Uranus over time (A-panels) also plots the value of the Hill sphere 
                  radius of Uranus (0.4469~au) as a red line. The time evolution of the value of the resonant angle, $\lambda_{\rm r}$ 
                  (B-panels), shows that this Trojan is transient. The variation of its semimajor axis, $a$, over time (C-panels) suggests 
                  that this object may remain as Uranian co-orbital for close to 1 Myr; the value of the semimajor axis of Uranus appears as 
                  a red line. The evolution of both eccentricity, $e$ (D-panels), and inclination, $i$ (E-panels), is far from regular. The
                  argument of perihelion, $\omega$ (F-panels), mostly circulates. The values of the distance to the nodes from the Sun 
                  (G-panels) oscillate somewhat regularly; the distance from the Sun to the descending (ascending) node is shown as a thick 
                  (dotted) line with Uranus' semimajor axis, and the distances to Saturn's aphelion and Neptune's perihelion, plotted as red 
                  lines.
                 }
         \label{controlyx49}
      \end{figure*}

        Fig. \ref{controlyx49}, central panels, shows the time evolution of various parameters for the nominal orbit of 2014~YX$_{49}$ in 
        Table \ref{elements} over an interval of $\pm$0.6 Myr. In addition to the nominal orbit, results for two other representative 
        control orbits are also displayed (left-hand and right-hand panels); they have been computed by adding (+) or subtracting ($-$) 
        6-times the uncertainty from the orbital elements (the six parameters) in Table \ref{elements}. The comparative evolution shows that 
        2014~YX$_{49}$ is a transient L$_4$ Trojan of Uranus. We have already pointed out above that close encounters of 2014~YX$_{49}$ with 
        Saturn or Neptune are not possible; Fig. \ref{controlyx49}, A-panels, indicate that no close approaches to Uranus under one Hill 
        radius are recorded during the studied time interval. Fig. \ref{controlyx49}, B-panels, shows that the duration of the present 
        Trojan episode of 2014~YX$_{49}$ is expected to be shorter than that of 2011~QF$_{99}$ (see fig. 4 in de la Fuente Marcos \& de la 
        Fuente Marcos 2014). In general, the dynamical evolution of 2014~YX$_{49}$ is more complex than that of 2011~QF$_{99}$ and multiple 
        transitions between the various incarnations of the 1:1 mean motion resonance with Uranus are observed. Our calculations show that
        2014~YX$_{49}$ may become a transient L$_5$ Trojan but also engage in quasi-satellite (the value of $\lambda_{\rm r}$ oscillates 
        around 0\degr) or horseshoe behaviour ($\lambda_{\rm r}$ librates around the Lagrangian point L$_3$, which is located at 180\degr{} 
        from Uranus on its orbit); trajectories hybrid of two (or more) elementary co-orbital states are also possible (see Fig. 
        \ref{controlyx49}, B-panels). During most of the Trojan phase, the amplitude of the oscillation of the tadpole motion is $<70\degr$.
%
%
%
%
        \hfil\par
        The evolution of the value of the semimajor axis (Fig. \ref{controlyx49}, C-panels) provides insight on how much time 2014~YX$_{49}$
        spends within Uranus' co-orbital zone. Our calculations indicate that this minor body may be co-orbital to Uranus for an average of 
        600 kyr with a minimum of about 400 kyr and a maximum close to 1 Myr. This makes it less long-term stable than both 2011~QF$_{99}$ 
        and 472651. In general, the values of $e$ (Fig. \ref{controlyx49}, D-panels) and $i$ (Fig. \ref{controlyx49}, E-panels) do not 
        exhibit obvious signs of coupling as those characteristic of the Kozai resonance (Kozai 1962); this is consistent with the behaviour 
        of the argument of perihelion (Fig. \ref{controlyx49}, F-panels) that mostly circulates. When compared with 2011~QF$_{99}$, 
        2014~YX$_{49}$ shows significantly less evidence of Kozai-like resonant behaviour. 
        \hfil\par
        Fig. \ref{controlyx49}, G-panels, shows the evolution of the distance to the nodes of 2014~YX$_{49}$ over time; the separation 
        between the Sun and the nodes has been computed by applying the expression $r = a (1 - e^2) / (1 \pm e \cos \omega)$, where the `+' 
        sign is used for the ascending node and the `$-$' sign for the descending node. In the case of Solar system objects that move in 
        orbits inclined with respect to the ecliptic plane, close encounters with major planets can only take place in the vicinity of the 
        nodes. Our calculations show that the evolution is quite regular and resembles that of 2011~QF$_{99}$ (see fig. 4, G-panels, in de 
        la Fuente Marcos \& de la Fuente Marcos 2014).
        \hfil\par
\begin{figure}
       \centering
        \includegraphics[width=\linewidth]{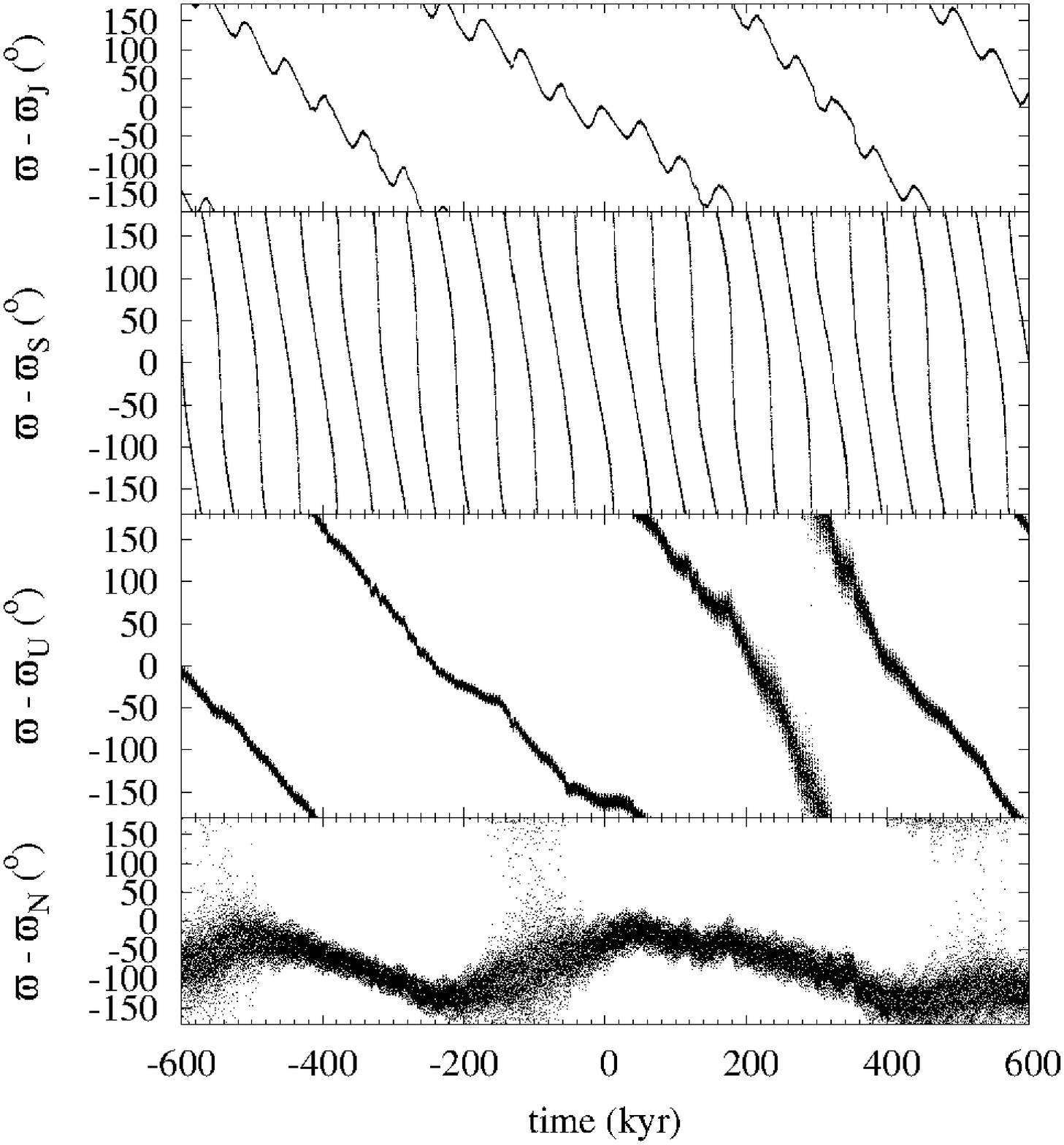}
        \caption{Evolution over time of the relative longitude of the perihelion, $\Delta \varpi$, of 2014~YX$_{49}$ with respect to the 
                 giant planets: referred to Jupiter ($\varpi - \varpi_{\rm J}$), to Saturn ($\varpi - \varpi_{\rm S}$), to Uranus ($\varpi -
                 \varpi_{\rm U}$), and to Neptune ($\varpi - \varpi_{\rm N}$). As in the case of (472651) 2015~DB$_{216}$, the relative 
                 longitudes do not librate or oscillate during the studied time window with the exception of that of Neptune. These results
                 are for the orbital solution in Table \ref{elements} although the observed behaviour is consistent across control orbits.
                }
        \label{secular}
     \end{figure}

        The secular dynamics of known Uranian co-orbitals has been studied in de la Fuente Marcos \& de la Fuente Marcos (2014) and de la 
        Fuente Marcos \& de la Fuente Marcos (2015). These studies singled out 472651 as peculiar because the precession frequency of the 
        longitude of the perihelion, $\varpi = \Omega + \omega$, of this object is not in secular resonance with Jupiter, Saturn or Uranus, 
        only with Neptune and for a limited amount of time. Fig. \ref{secular} shows the time evolution of the relative longitude of the 
        perihelion, $\Delta \varpi$, of 2014~YX$_{49}$ with respect to the giant planets. The precession frequency of the longitude of the 
        perihelion of 2014~YX$_{49}$ is only in secular resonance with Neptune, but for the entire simulation. The value of $\Delta \varpi = 
        \varpi - \varpi_{\rm N}$ librates around -90\degr{} (or 270\degr). This type of secular behaviour is not observed for any other 
        known Uranian co-orbital. The time evolution of the relative longitude of the perihelion with respect to Jupiter, Saturn and Uranus 
        closely resembles that of 472651 (see fig. 5 in de la Fuente Marcos \& de la Fuente Marcos 2015).
%
%
%
%

     \subsection{Stability analysis}
        The overall dynamical stability associated with the path currently followed by 2014~YX$_{49}$ has been discussed in Sections 3.1 and 
        3.2; the orbit of 2014~YX$_{49}$ is not long-term stable even if no truly close encounters with planetary bodies are observed. In 
        absence of sufficiently close approaches to Uranus, the dynamical evolution of this minor body must be fully driven by mean motion 
        resonances with the giant planets. Here, we investigate what sequence of resonant events placed 2014~YX$_{49}$ in its present orbit 
        and how was it temporarily stabilized. We also explore the circumstances that will surround its future ejection from the Trojan 
        state.
        \hfil\par
        As pointed out by de la Fuente Marcos \& de la Fuente Marcos (2014), the Uranian Trojan 2011~QF$_{99}$ is not currently engaged in 
        resonant behaviour with Jupiter or Neptune, but it is subjected to the 7:20 mean motion resonance with Saturn. This dynamical 
        response is characteristic of Uranian Trojans and it was first predicted by Gallardo (2006). Although the shape and orientation of 
        the orbits of 2011~QF$_{99}$ and 2014~YX$_{49}$ are rather different, the perturbational scenario that drives their evolution is 
        virtually identical and their resonant profiles must also be very similar. In addition to the 1:1 mean motion resonance with Uranus 
        in any of its forms (elementary or composite), the three mean motion resonances of interest here are 1:7 with Jupiter, 7:20 with 
        Saturn and 2:1 with Neptune. Their respective resonant arguments are $\sigma_{\rm J} = 7 \lambda - \lambda_{\rm J} - 6 \varpi$, 
        $\sigma_{\rm S} = 20 \lambda - 7 \lambda_{\rm S} - 13 \varpi$ and $\sigma_{\rm N} = \lambda - 2 \lambda_{\rm N} + \varpi$, where 
        $\lambda_{\rm J}$, $\lambda_{\rm S}$ and $\lambda_{\rm N}$ are the respective mean longitudes of Jupiter, Saturn and Neptune. 
        \hfil\par
\begin{figure}
        \centering
         \includegraphics[width=\linewidth]{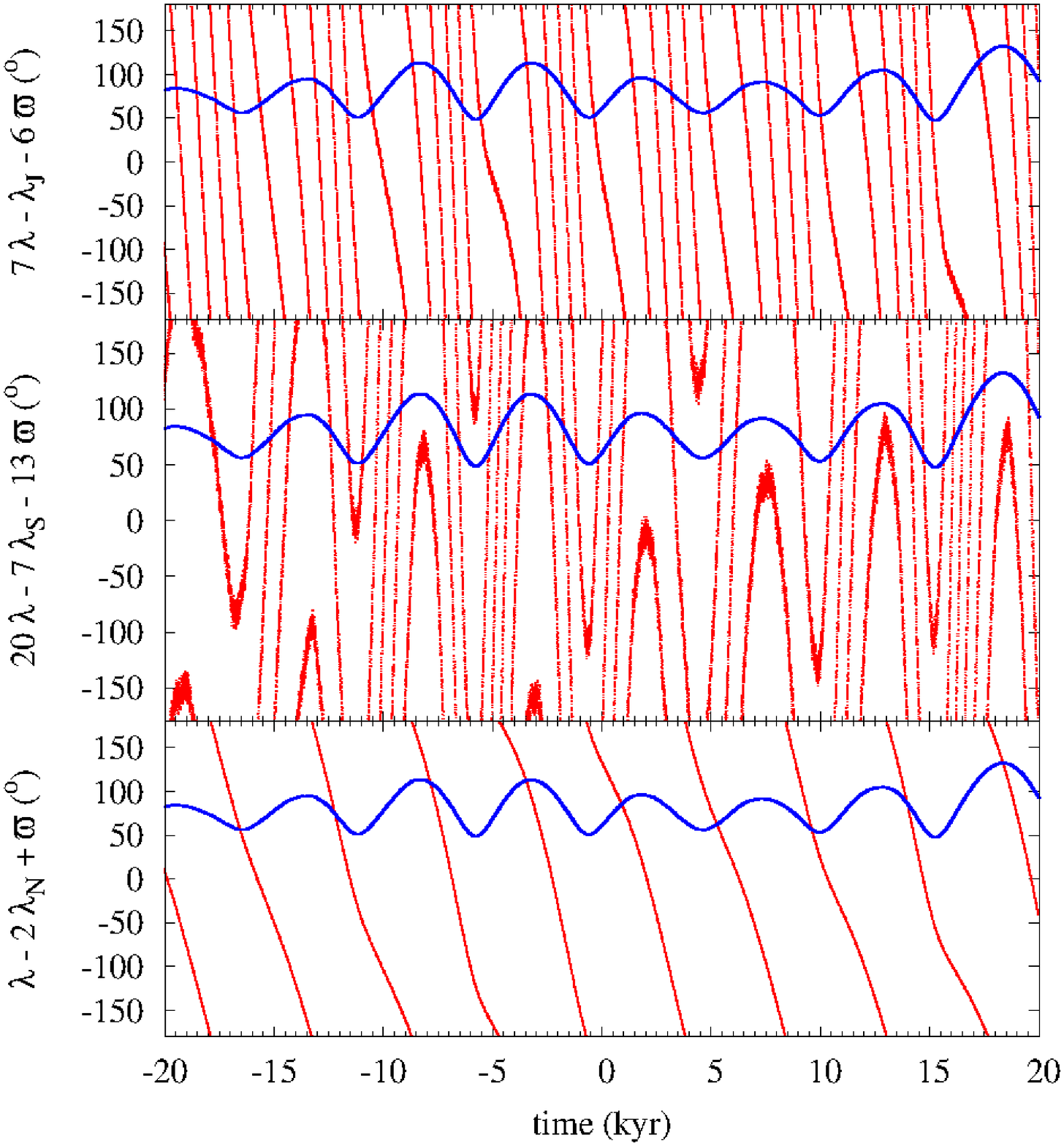}
         \caption{Variation of the various resonant arguments of interest in the analysis of the stability of 2014~YX$_{49}$ for the orbital 
                  determination in Table \ref{elements}. Time evolution of $\sigma_{\rm J} = 7 \lambda - \lambda_{\rm J} - 6 \varpi$ (top 
                  panel), $\sigma_{\rm S} = 20 \lambda - 7 \lambda_{\rm S} - 13 \varpi$ (middle panel) and $\sigma_{\rm N} = \lambda - 2 
                  \lambda_{\rm N} + \varpi$ (bottom panel) for the interval (-20, 20) kyr; the thick line shows the behaviour of the 
                  relative mean longitude with respect to Uranus, same data as in Fig. \ref{controlyx49}, central B-panel. The resonant 
                  evolution depicted here is consistent across control orbits.
                 }
         \label{resonances}
      \end{figure}

        Fig. \ref{resonances} shows the time evolution of the various resonant arguments pointed out above. It is fully consistent with the 
        behaviour depicted in fig. 5 of de la Fuente Marcos \& de la Fuente Marcos (2014) for 2011~QF$_{99}$ and it confirms that, during
        its current Trojan episode, 2014~YX$_{49}$ is simultaneously trapped in the 7:20 mean motion resonance with Saturn. The resonant 
        argument $\sigma_{\rm S}$ alternates between circulation and asymmetric libration because the motion is chaotic. During the Trojan 
        resonant episode, 2014~YX$_{49}$ is subjected to a three-body resonance: the 1:1 mean motion resonance with Uranus and the 7:20 mean 
        motion resonance with Saturn.
%
%
%
%
        \hfil\par
\begin{figure*}
        \centering
         \includegraphics[width=0.33\linewidth]{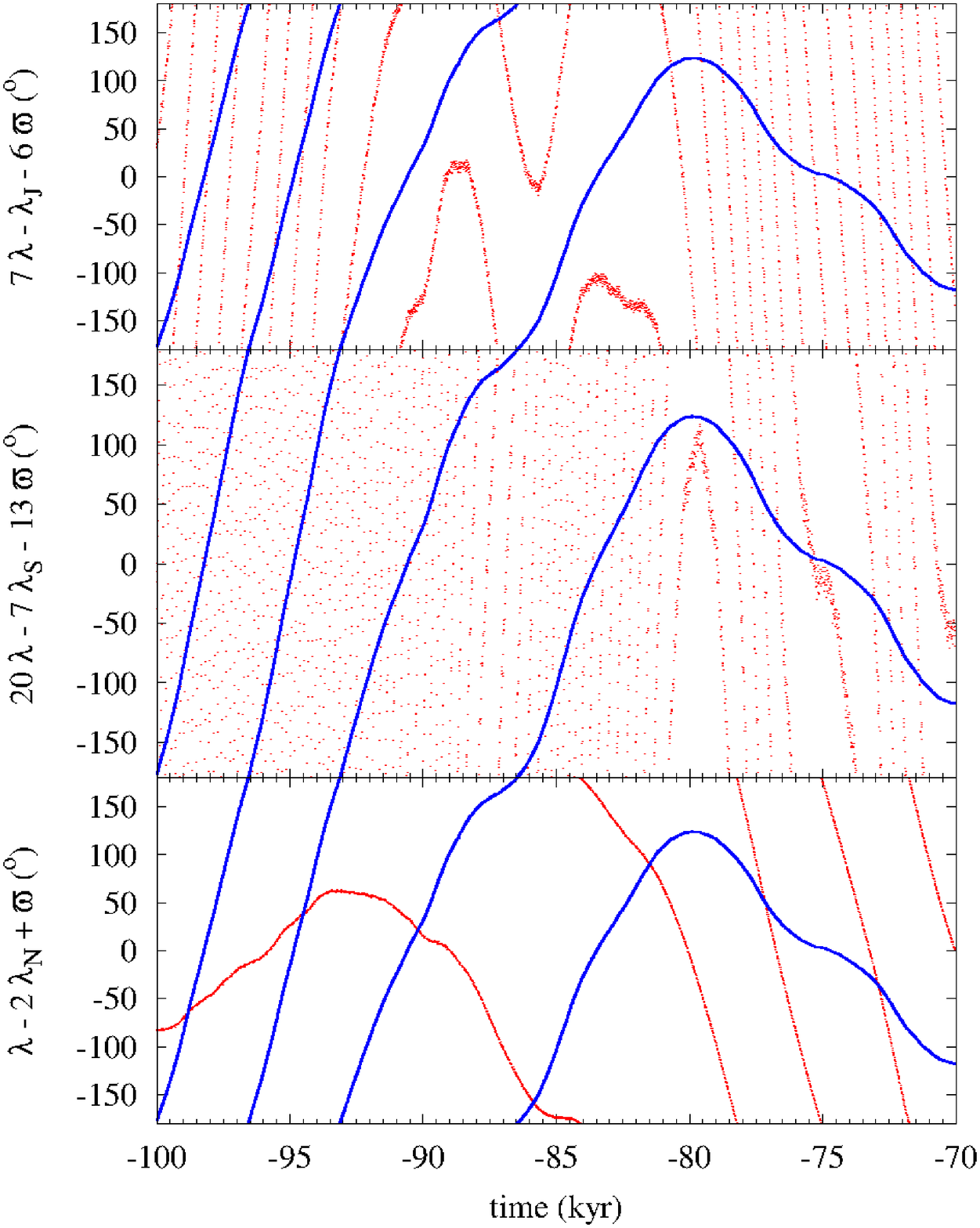}
         \includegraphics[width=0.33\linewidth]{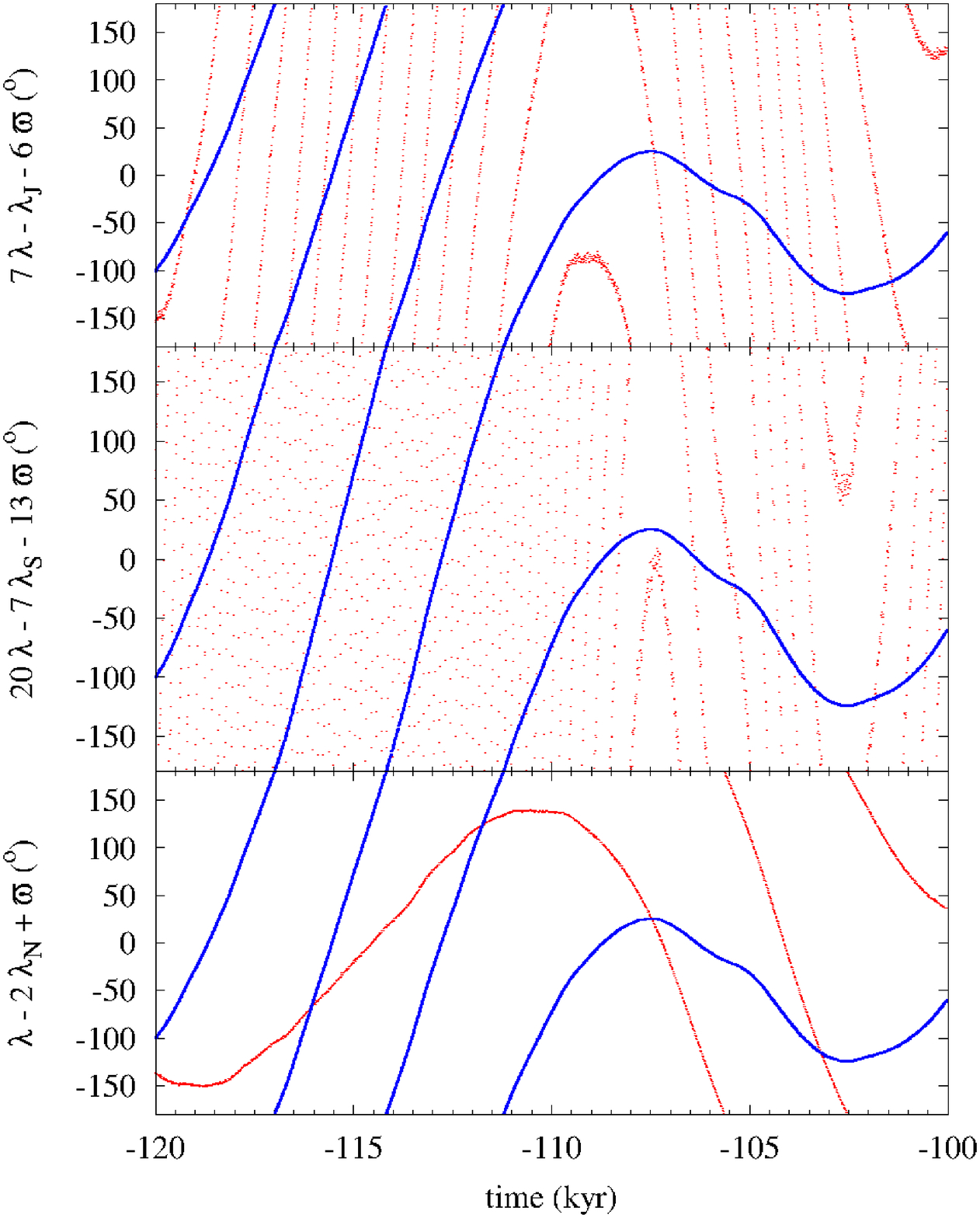}
         \includegraphics[width=0.33\linewidth]{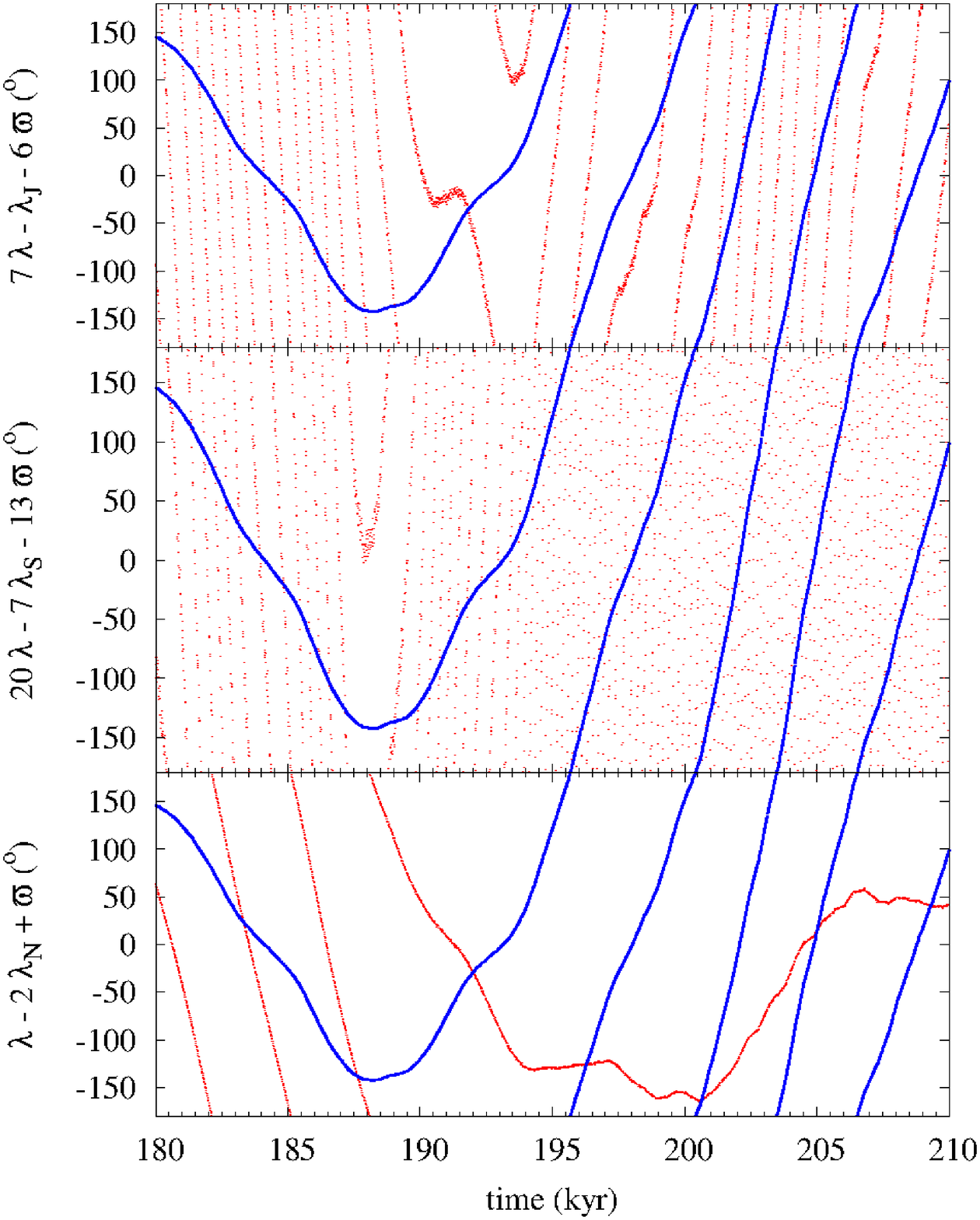}
         \caption{Two examples of capture into the 1:1 mean motion resonance with Uranus for two different control orbits of 2014~YX$_{49}$ 
                  (left-hand and central panels) and one example of ejection (right-hand panels). 
                  Resonant arguments as in Fig. \ref{resonances}.
                 }
         \label{examples}
      \end{figure*}

        In de la Fuente Marcos \& de la Fuente Marcos (2014), we showed that 2011~QF$_{99}$ was captured into Uranus' co-orbital zone after
        experiencing a three-body resonance with Jupiter and Neptune. Fig. \ref{examples} shows that the same mechanism drives the insertion 
        of 2014~YX$_{49}$ in its present resonant state (see left-hand and central panels). As Newtonian gravitation is time symmetric, the 
        mechanism leading to a certain resonant state should also be able to take the affected minor body out of it. This was already 
        discussed in de la Fuente Marcos \& de la Fuente Marcos (2014) for the case of 2011~QF$_{99}$ and Fig. \ref{examples}, right-hand 
        panels, shows that it is also certain for 2014~YX$_{49}$. However, the details of the resonant events behind the transitions between  
        co-orbital states are different although they involve ephemeral or very brief multibody mean motion resonances.
%
%
%
%
        \hfil\par
\begin{figure*}
        \centering
         \includegraphics[width=0.49\linewidth]{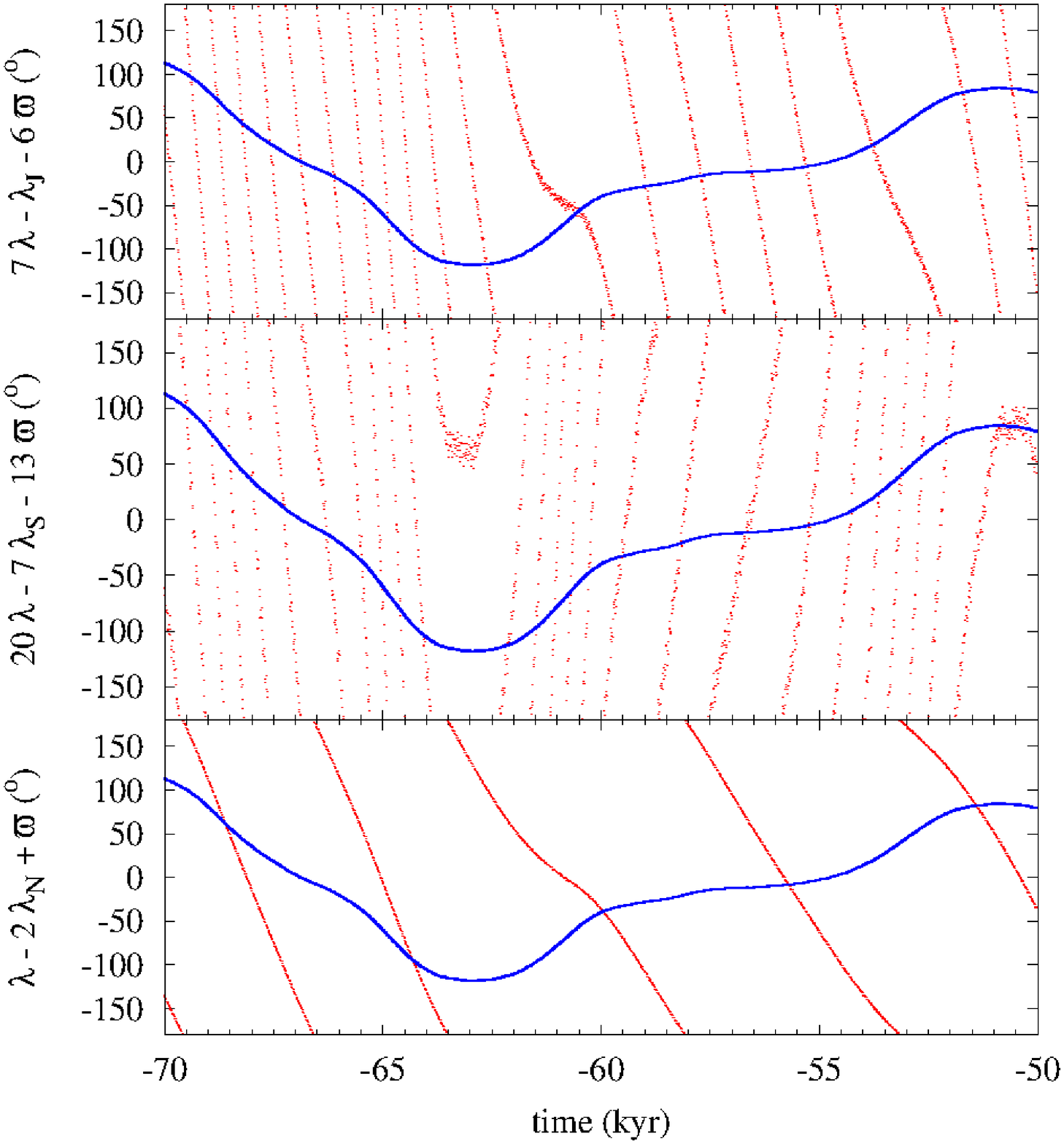}
         \includegraphics[width=0.49\linewidth]{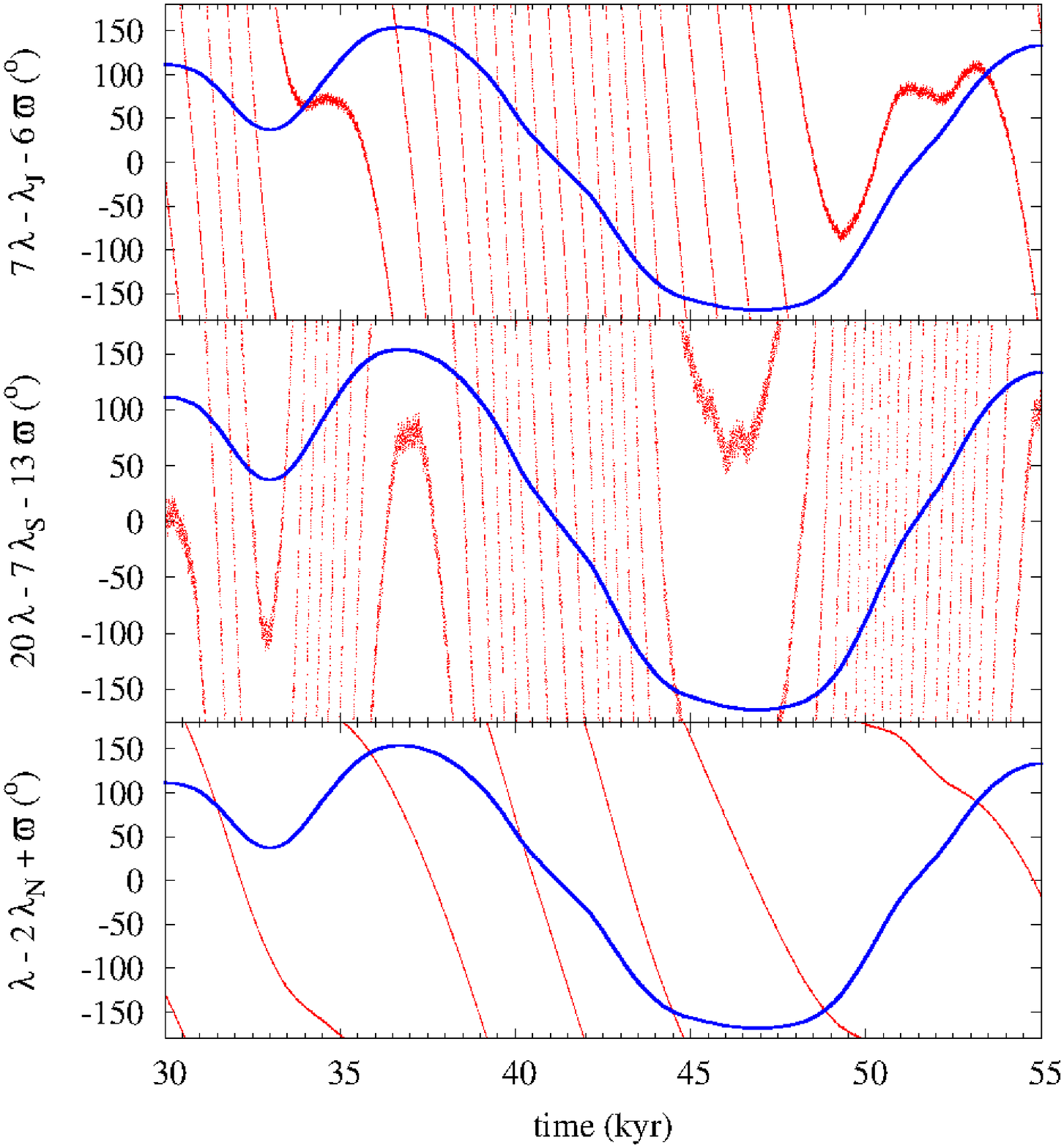}
         \caption{Two examples of transitions of 2014~YX$_{49}$ within the 1:1 mean motion resonance with Uranus induced by very brief
                  resonant episodes with Jupiter. During these ephemeral events, a three-body resonance with Jupiter and Uranus is active.
                  Resonant arguments as in Fig. \ref{resonances}.
                 }
         \label{transition}
      \end{figure*}

        Becoming a jumping Trojan, i.e. going from L$_4$ to L$_5$ and (perhaps) back, also involves three-body resonances but in this case
        with Jupiter and Uranus. Fig. \ref{transition} shows two examples of this type of behaviour. However, Neptune may also be involved
        when transitioning to the horseshoe state (de la Fuente Marcos \& de la Fuente Marcos 2014). Additional examples of similar 
        transitions for the cases of 2011~QF$_{99}$ and 472651 can be found in de la Fuente Marcos \& de la Fuente Marcos (2014) and de la 
        Fuente Marcos \& de la Fuente Marcos (2015), respectively. As pointed out in de la Fuente Marcos \& de la Fuente Marcos (2015), the
        outcome of these rapid transitions is unusually sensitive to the physical model used to perform the calculations. A five-body 
        physical model including the Sun and the four outer planets such as the one used by Marzari et al. (2003) or Alexandersen et al. 
        (2013b) may not be able to arrive to realistic conclusions regarding the resonant details associated with the transitions between
        co-orbital states experienced by objects like 2011~QF$_{99}$, 2014~YX$_{49}$ and 472651.
%
%
%
%

  \section{Discussion}
     The discovery of 2014~YX$_{49}$, the second Uranian Trojan, confirms that, contrary to the conclusions reached by Horner \& Evans 
     (2006), Uranus can capture efficiently minor bodies into the 1:1 commensurability. In addition, the fact that it was originally found 
     at a declination above +30\degr{} by Pan-STARRS strongly suggests that the number of objects moving in similar orbits may not be 
     negligible (see the detailed discussion in section 8 of de la Fuente Marcos \& de la Fuente Marcos 2014). Asteroid (472651) 
     2015~DB$_{216}$ is another example of transient Uranian co-orbital discovered under similar circumstances, i.e. serendipitously. If the 
     analysis in de la Fuente Marcos \& de la Fuente Marcos (2015) showed that there is enough theoretical ground to assume that a 
     population of high orbital inclination Uranian co-orbitals may exist, the identification of this Trojan shows that there are also  
     robust observational grounds, adding more weight to the hypothesis. Both 2014~YX$_{49}$ and 472651 are comparatively large and bright,
     making them easier to detect; finding any fainter, hypothetical bodies moving in somewhat similar orbits may require a carefully 
     planned survey though.
     \hfil\par
     Dvorak, Bazs\'o \& Zhou (2010) investigated the stability of putative primordial Uranus' Trojans and found stable islands at the 
     inclination intervals (0, 7)\degr, (9, 13)\degr, (31, 36)\degr and (38, 50)\degr. Asteroid 2011~QF$_{99}$, the first Uranian Trojan, 
     inhabits one of these stable areas as its current value of the orbital inclination is nearly 11\degr{}; however, this object is not a 
     primordial Uranus' Trojan (Alexandersen et al. 2013b; de la Fuente Marcos \& de la Fuente Marcos 2014). In contrast, 2014~YX$_{49}$ 
     currently occupies (see Fig. \ref{controlyx49}, E-panels) an unstable region between two of the stable islands described in Dvorak et 
     al. (2010) as its current value of the inclination is nearly 26\degr. In addition, its inclination remains in the range 20--27\degr{} 
     for most of the studied time. However, our analysis shows that even if 2014~YX$_{49}$ is less dynamically stable than 2011~QF$_{99}$, 
     it can still remain as transient Trojan of Uranus for a relatively long period of time. The situation is similar to that of 472651, 
     although this object moves in a higher inclination orbit.
     \hfil\par
     Our results provide additional confirmation that overlapping mean motion resonances can trigger transitions between co-orbital states 
     and also deliver minor bodies inside the co-orbital zone of a planetary body as well as evict them from there. The duration of these 
     episodes is short, typically 10--20~kyr in the cases of captures as co-orbitals or ejections from the co-orbital zone, and about 
     5~kyr when inducing transitions between the various co-orbital states (see Figs \ref{examples} and \ref{transition}). The dynamical 
     effects of three-body mean motion resonances have been studied recently by Gallardo (2014).

  \section{Conclusions}
     In this paper, we provide a detailed exploration of the past, present and future orbital evolution of 2014~YX$_{49}$, the second
     Trojan of Uranus. Our conclusions can be summarized as follows.
     \begin{enumerate}[(i)]
        \item Asteroid 2014~YX$_{49}$ is a transient L$_4$ Trojan of Uranus, the second minor body to be confirmed as currently engaged in 
              this type of resonant state. Its current libration period is over 5\,100~yr.
        \item Although 2014~YX$_{49}$ may have remained in its current Trojan state for about 60~kyr, it can continue doing so for another 
              80~kyr. Our calculations suggest that it may stay within Uranus' co-orbital zone for nearly 1~Myr. 
        \item In addition to being a Uranian Trojan, 2014~YX$_{49}$ is currently trapped in the 7:20 mean motion resonance with Saturn;
              therefore, this minor body is currently subjected to a three-body resonance. It is also engaged in a long-term secular 
              resonance with Neptune.
        \item The dynamical mechanism behind the insertion of 2014~YX$_{49}$ in Uranus' co-orbital zone involves a brief episode of 
              concurrent resonant behaviour with Jupiter (1:7) and Neptune (2:1). A virtually identical mechanism is also responsible for 
              scattering 2014~YX$_{49}$ away from the 1:1 commensurability with Uranus.   
        \item Transitions between the various states within the 1:1 mean motion resonance with Uranus are mainly triggered by ephemeral 
              resonant episodes with Jupiter.             
        \item Our analysis shows that 2014~YX$_{49}$ shares a number of dynamical characteristics with (472651) 2015~DB$_{216}$, another 
              Uranian co-orbital that follows a relatively high-inclination orbit. As these two objects are comparatively rather 
              bright/large, the presence of fainter/smaller minor bodies following somewhat similar paths is highly probable and this 
              plausible conjecture lends credibility to the idea that a population of temporary Uranian co-orbitals may exist at relatively 
              high orbital inclinations.
     \end{enumerate}
     Our investigation of the dynamics of 2011~QF$_{99}$, 472651 and 2014~YX$_{49}$ shows that ephemeral multibody mean motion resonance 
     episodes are behind the events that lead to the capture and ejection of transient Uranian co-orbitals. This general mechanism may have 
     also played some role in the loss of any putative primordial Uranus' Trojans early in the history of the Solar system. 

  \section*{Acknowledgements}
     We thank the anonymous referee for a prompt report, S.~J. Aarseth for providing the code used in this research, A. I. G\'omez de 
     Castro, I. Lizasoain and L. Hern\'andez Y\'a\~nez of the Universidad Complutense de Madrid (UCM) for providing access to computing 
     facilities, and S. Deen for finding precovery images of 2014~YX$_{49}$ that improved the orbital solution of this object and for 
     additional comments. This work was partially supported by the Spanish `Ministerio de Econom\'{\i}a y Competitividad' (MINECO) under 
     grant ESP2014-54243-R. Part of the calculations and the data analysis were completed on the EOLO cluster of the UCM, and we thank S. 
     Cano Als\'ua for his help during this stage. EOLO, the HPC of Climate Change of the International Campus of Excellence of Moncloa, is 
     funded by the MECD and MICINN. This is a contribution to the CEI Moncloa. In preparation of this paper, we made use of the NASA 
     Astrophysics Data System, the ASTRO-PH e-print server, and the MPC data server.

  \bsp
  \label{lastpage}
\end{document}